\documentclass[prd, aps, floatfix, twocolumn, superscriptaddress, preprintnumbers]{revtex4-1}

\usepackage[colorlinks = true, citecolor = blue, linkcolor = blue, urlcolor = blue]{hyperref}

\usepackage{graphicx}
\usepackage{tabularx}
\usepackage[normalem]{ulem}

\usepackage{mathtools}
\usepackage{amsmath}
\usepackage[capitalise]{cleveref} %% automatically format reference notations, has to be loaded after 'amsmath'
\usepackage{amssymb}
\usepackage{bm}
\usepackage{mathrsfs}
\usepackage{slashed}
\usepackage{physics}
\usepackage{xcolor}
\usepackage{diagbox}
\newcommand{\nnu}{\nonumber\\}

\newcommand{\Qs}{Q_s}
\newcommand{\rt}{\boldsymbol{r}}
\newcommand{\bt}{\boldsymbol{b}}

\usepackage{atbegshi,picture}
\AtBeginShipoutNext{\AtBeginShipoutUpperLeft{%
\put(\dimexpr\paperwidth-2.75cm\relax,-1.1cm){\makebox[0pt][r]{MIT-CTP/5649}}%
}}
\begin{document}
\title{Transverse Energy-Energy Correlators in the Color-Glass Condensate at the Electron-Ion Collider}

\author{Zhong-Bo Kang}
\email{zkang@ucla.edu}
\affiliation{Department of Physics and Astronomy, University of California, Los Angeles, CA 90095, USA}
\affiliation{Mani L. Bhaumik Institute for Theoretical Physics, University of California, Los Angeles, CA 90095, USA}
\affiliation{Center for Frontiers in Nuclear Science, Stony Brook University, Stony Brook, NY 11794, USA}

\author{Jani Penttala}
\email{janipenttala@physics.ucla.edu}
\affiliation{Department of Physics and Astronomy, University of California, Los Angeles, CA 90095, USA}
\affiliation{Mani L. Bhaumik Institute for Theoretical Physics, University of California, Los Angeles, CA 90095, USA}

\author{Fanyi Zhao}
\email{fanyi@mit.edu}
\affiliation{Center for Theoretical Physics, Massachusetts Institute of Technology, Cambridge, MA 02139, USA}

\author{Yiyu Zhou}
\email{zyiyu@m.scnu.edu.cn}
\affiliation{Key Laboratory of Atomic and Subatomic Structure and Quantum Control (MOE), Guangdong Basic Research Center of Excellence for Structure and Fundamental Interactions of Matter, Institute of Quantum Matter, South China Normal University, Guangzhou 510006, China}
\affiliation{Guangdong-Hong Kong Joint Laboratory of Quantum Matter, Guangdong Provincial Key Laboratory of Nuclear Science, Southern Nuclear Science Computing Center, South China Normal University, Guangzhou 510006, China }
\affiliation{Department of Physics and Astronomy, University of California, Los Angeles, CA 90095, USA}

\begin{abstract}
We investigate the transverse energy-energy correlators (TEEC) in the small-$x$ regime at the upcoming Electron-Ion Collider (EIC). Focusing on the back-to-back production of electron-hadron pairs in both $ep$ and $eA$ collisions, we establish a factorization theorem given in terms of the hard function, quark distributions, soft functions, and TEEC jet functions, where the gluon saturation effect is incorporated. Numerical results for TEEC in both $ep$ and $eA$ collisions are presented, together with the nuclear modification factor $R_A$. Our analysis reveals that TEEC observables in deep inelastic scattering provide a valuable approach for probing gluon saturation phenomena. Our findings underscore the significance of measuring TEEC at the EIC, emphasizing its efficacy in advancing our understanding of gluon saturation and nuclear modifications in high-energy collisions.

\end{abstract}

\maketitle

\section{Introduction}
Event shape observables, crucial for understanding energy flow and correlations in high-energy scattering processes, have been extensively explored in various collision scenarios~\cite{SLD:1994idb,L3:1992btq,OPAL:1991uui,TOPAZ:1989yod,TASSO:1987mcs,JADE:1984taa,Fernandez:1984db,Wood:1987uf,CELLO:1982rca,PLUTO:1985yzc,OPAL:1990reb,ALEPH:1990vew,L3:1991qlf,SLD:1994yoe,ATLAS:2015yaa,ATLAS:2017qir,ATLAS:2020mee} such as $pp$, $ep$, $e^+e^-$, and others. These studies shed light on different dynamical properties of Quantum Chromodynamics (QCD). The event shape observables play a significant role not only in determining the strong coupling constant $\alpha_s$ and verifying asymptotic freedom but also in refining non-perturbative QCD power corrections and probing potential new physics phenomena. Especially, there exists an opportunity to study these observables theoretically and compare them with experimental measurements for the deep-inelastic scattering (DIS) processes at the upcoming Electron-Ion Collider (EIC)~\cite{Boer:2011fh, Accardi:2012qut, AbdulKhalek:2021gbh, AbdulKhalek:2022hcn}.

Numerous endeavors are dedicated toward the investigation of event-shape observables within the context of DIS. In this context, our focus is directed towards the Transverse Energy-Energy Correlation (TEEC) event shape observable in DIS. TEEC, as introduced in~\cite{Ali:1984yp}, originates as an extension of the Energy-Energy Correlation (EEC)~\cite{Basham:1978bw,Basham:1978zq} that was introduced for $e^+e^-$ collisions to characterize global event shapes. In the environment of hadronic colliders, the event shape observable can be extended by considering the transverse energy of the hadrons~\cite{Ali:2012rn, Gao:2019ojf}.
In the realm of DIS, the generalization of TEEC occurs through the application of the transverse energy correlation between the lepton and hadrons in the final state in the {\it lab frame} of lepton-proton collisions, which was initially conducted in Ref.~\cite{Li:2020bub}.
As demonstrated in Ref.~\cite{Li:2020bub}, with the angle $\phi$ defined as the azimuthal angle difference between the produced electron and hadron transverse momentum, resummed predictions in the limit of back-to-back $\phi\rightarrow\pi$ configurations can be obtained with high accuracy, allowing for reliable calculations of the distribution of $\phi$ across the entire range of $[0,\pi]$.
EEC and TEEC present a notable advantage in that the contribution from soft radiation is effectively suppressed due to its low energy.
Consequently, the impact of hadronization effects is anticipated to be comparatively small when contrasted with other event-shape observables. Another advantage of the TEEC lies in the fact that the collision kinematics can be accurately reconstructed in the {\it lab frame} as pointed out in Ref.~\cite{Gao:2022bzi}, and thus the TEEC can serve as great probes for the transverse-momentum dependent structures of the proton~\cite{Li:2020bub,Kang:2023big}. In DIS, TEEC also offers a precise approach for determining the strong coupling, like the analyses in Refs.~\cite{Catani:1996vz,Graudenz:1997gv,Nagy:2001xb}, and facilitates the exploration of nuclear dynamics as discussed in Refs.~\cite{Kang:2013wca,Kang:2013lga}. 

On the other hand, it has long been realized that the extracted parton distribution functions (PDFs) from experimental data, particularly the gluon distribution, exhibit a rapid increase as the partonic longitudinal momentum fraction, $x$, diminishes. The evolution of the gluon density at high energies, under the condition of fixed momentum transfer $Q^2$, is encapsulated by the Balitsky-Fadin-Kuraev-Lipatov (BFKL) evolution equation~\cite{Lipatov:1976zz,Kuraev:1977fs,Balitsky:1978ic}. The BFKL equation, a linear evolution equation, describes the evolution of the gluon distribution in terms of $x$. Its solution manifests a sharp increase as $x$ decreases. Nonetheless, the gluon density is constrained from escalating indefinitely at high energies.
In experimental observations, compelling evidence has emerged, especially at diminutive $x$ values, indicating the presence of a distinct QCD regime known as the saturation regime.
This regime eludes comprehensive explication through conventional linear QCD evolution frameworks~\cite{Gribov:1983ivg,Stasto:2000er,Armesto:2004ud,Gelis:2010nm}.

Searching for the gluon saturation phenomenon~\cite{Gribov:1983ivg,Mueller:1985wy,Mueller:1989st,Dumitru:2017cwt,McLerran:1993ka,McLerran:1994vd,Iancu:2003xm,Gelis:2010nm} is one of the scientific goals of the future EIC.
The saturation physics refers to a phenomenon where the gluon density becomes so dominating that the interactions among gluons become significant, leading to a saturation of parton densities at small values of the partonic longitudinal momentum fraction $x$. Namely, this saturation occurs at high energy and small $x$, characterized by a saturation scale, denoted as $Q_s$. Traditional linear QCD evolution equations, such as the BFKL equation, no longer accurately describe the dynamics in this regime~\cite{Gribov:1983ivg,Mueller:1985wy}. One then needs the non-linear extension of the BFKL equation, the Balitsky-Kovchegov (BK) equation~\cite{Balitsky:1995ub,Kovchegov:1999yj}. This non-linear dynamic phenomenon can be characterized better when a nuclear target is involved, wherein the interaction extends across a longitudinal distance approximately equal to or greater than the size of the nucleus. Under these conditions, the individual nucleons positioned at the same impact parameter become indistinguishable.
Gluons originating from distinct nucleons have the potential to magnify the overall transverse gluon density by a factor of $A^{1/3}$ with $A$ being the mass number of the target. Therefore, a substantial alteration in the TEEC is expected when the target hadron is substituted from a proton to a heavy nucleus like gold. Consequently, this novel observable, when explored at the forthcoming EIC, has the potential to provide further compelling evidence for parton saturation.

The rest of the paper is structured as follows.
\Cref{sec:th} provides the theoretical formalism for TEEC in DIS. We explain each component in the factorization, including the quark distribution in the small-$x$ region and a detailed construction of the TEEC jet function.
\Cref{sec:pheno} presents our phenomenological study to demonstrate the potential of TEEC observables for probing gluon saturation and nuclear modification effects using $ep/eA$ collisions.
Finally, we conclude our work in \cref{sec:conclusion}.

\section{Theoretical formalism}
\label{sec:th}
In this section, following the theoretical formalism of TEEC in Deep Inelastic Scattering \cite{Li:2020bub}, we study the transverse energy-energy correlation between the lepton and hadrons in the final state:
\begin{align}
e(\ell)+p/A(P_A) \to e(\ell') + h(P_h) + X
\, ,  
\end{align}
where the scattered electron and final hadron are produced in a back-to-back configuration in the transverse plane.
\begin{figure*}[htb]
\centering
\includegraphics[width = 0.8 \textwidth]{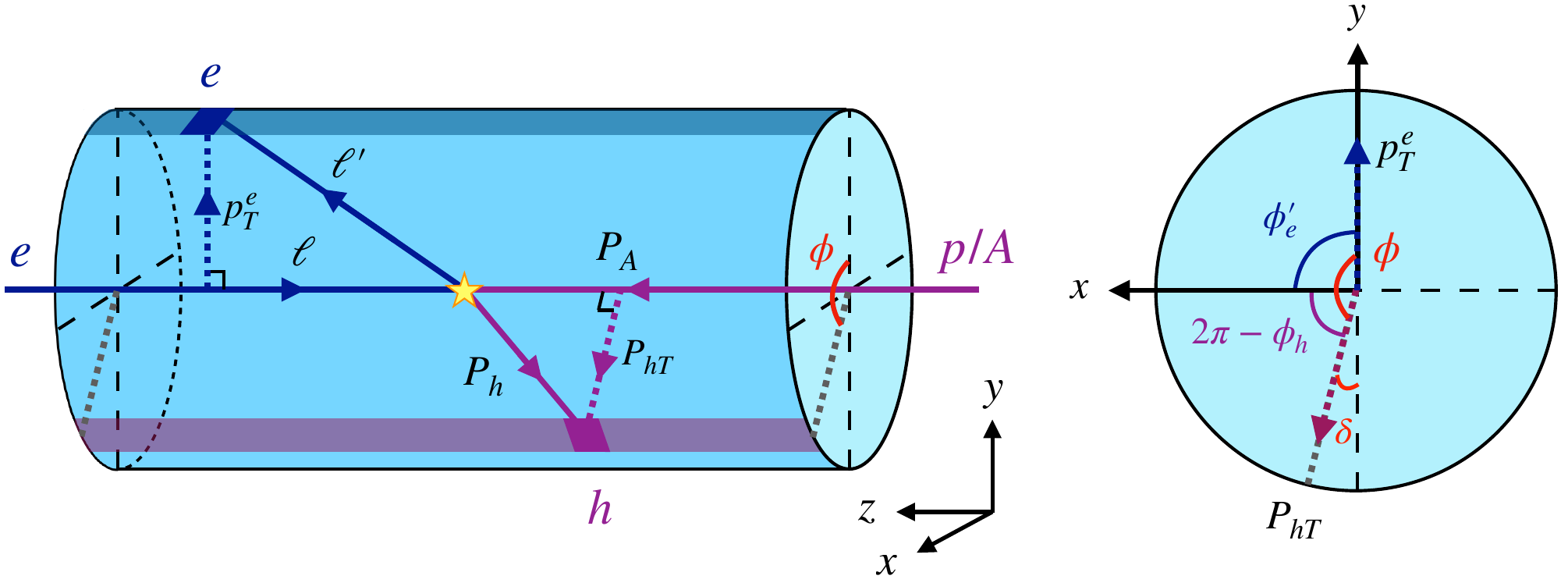}
\caption{Illustration of TEEC for DIS in the lab frame (left panel).
The incoming proton momentum $P_A$ and electron momentum $\ell$ define the $z$-axis.
We align the transverse momentum of the outgoing electron $\boldsymbol{p}_T^e$ with the $+y$-direction to define the $xy$-plane (right panel).}
\label{f.reaction}
\end{figure*}
The TEEC is illustrated in \cref{f.reaction} and defined as:
\begin{align}
\allowdisplaybreaks
\textbf{TEEC}
& =
\sum_{h} \int \dd{\sigma_{\mathrm{DIS}}}
\frac{E_{T,l} E_{T,h}}{E_{T,l} \sum_{i} E_{T,i}}
\delta \pqty{\tau-\frac{1+\cos{\phi}}{2}} 
\nnu
& =
\sum_{h} \int \dd{\sigma_{\mathrm{DIS}}}
\frac{E_{T,h}}{\sum_{i} E_{T,i}}
\delta \pqty{\tau-\frac{1+\cos\phi}{2}}
\, ,  
\label{e.teec}
\end{align}
where the sum runs over all the hadrons in the final state, and we define the variable $\tau$ as:
\begin{align}
\tau = \frac{1 + \cos{\phi}}{2}\, .
\end{align}
Here $\phi$ is the azimuthal angle between the final-state lepton $e$ and hadron $h$ as shown in the right panel of \cref{f.reaction}.
We have also defined the angle $\delta = \pi-\phi = \pi-(2\pi-\phi_h+\phi_e') = \phi_h-\phi_e'-\pi$, which is a small angle under the back-to-back limit, $\phi \to \pi$. Correspondingly, we have $\tau\ll 1$. As we have mentioned in the Introduction, we analyze the event in the center-of-mass frame of the lepton and proton collisions, with the proton (or the nucleus) moving in the $+z$ direction while the incoming lepton moving in the $-z$ direction, as shown in~\cref{f.reaction}. 

The TMD factorization theorem for the TEEC observable in the back-to-back region (i.e. $\tau\ll 1$) is given by~\cite{Li:2020bub,Gao:2023ulg}:
\begin{widetext}
\begin{align}
\textbf{TEEC} \equiv \frac{\dd{\sigma}}{\dd{\tau} \dd{y_e} \dd[2]{\boldsymbol{p}_T^e}}
= &\,
\sigma_0 H \pqty{Q, \mu}
\sum_{q} e_{q}^2 \frac{p_T^e}{\sqrt{\tau}}
\int_{-\infty}^{\infty} \frac{\dd{b}}{2 \pi} e^{-2 i b \sqrt{\tau} p_T^e}   
f_{q}^{(u)} \pqty{x, b, \mu, \zeta/\nu^2}
S_{nn_h} \pqty{b, \mu, \nu}
J_{q}^{(u)} (b, \mu, \zeta'/\nu^2)
\label{e.factorization_unsubtracted}
\\
=&\,\sigma_0 H \pqty{Q, \mu}
\sum_{q} e_{q}^2 \frac{p_T^e}{\sqrt{\tau}}
 \int_0^\infty \frac{\dd{b}}{\pi} \cos(2b\sqrt{\tau} p_T^e)f_{q}^{(u)} \pqty{x, b, \mu, \zeta/\nu^2}
S_{nn_h} \pqty{b, \mu, \nu}
J_{q}^{(u)} (b, \mu, \zeta'/\nu^2)
\, .
\nonumber
\end{align}
\end{widetext}
Even though the TEEC is the cross section weighted by the hadron momentum fraction as in~\cref{e.teec}, we abuse the notation a bit by still denoting it as $\dd \sigma$. Here $y_e$ and $\boldsymbol{p}_T^e$ are the rapidity and transverse momentum of the produced lepton in the laboratory frame with respect to the beam direction, and we take the outgoing lepton to lie along the $y$-axis. On the other hand, $f_{q}^{(u)} \pqty{x, b, \mu, \zeta/\nu^2}$ is the ``unsubtracted'' TMD quark distribution, where $b$ is the $x$-component of the $\bm{b}$ vector in the standard quark TMD distribution as probed e.g. in semi-inclusive DIS~\cite{Boussarie:2023izj,Bacchetta:2006tn}. In other words, we have $\bm{b}\equiv(b_x, b_y)=(b, 0)$ and thus the integration limits are given by $b\in(-\infty, \infty)$ in the first line of~\cref{e.factorization_unsubtracted}. It is important to realize that the cross section is differential in variable $\tau$ (i.e. azimuthal angle $\phi$), which is related to the $x$ component of the transverse momentum of the final observed hadron,
\begin{align}
    |P_{hx}|/z = P_{hT}/z|\sin\delta| \approx \, 2 \sqrt{\tau}\,p_T^e\,,
\end{align}
where $z$ is the momentum fraction of the quark carried by the hadron fragmenting from it. Consequently, we have a one-dimensional Fourier transform, i.e. only the $x$ component of the conjugated coordinate variable $\bm{b}$ is relevant. This has been derived clearly in~\cite{Fang:2023thw,Gao:2023ulg,Li:2020bub}. $S_{nn_h} \pqty{b, \mu, \nu}$ is the soft function representing the contribution from soft gluon radiation, and $H(Q, \mu)$ is the hard function. At the same time, $J_{q}^{(u)} (b, \mu, \zeta'/\nu^2)$ is the ``unsubtracted'' TEEC jet function, which has a close relation with the TMD fragmentation functions as given below. On the second line of~\cref{e.factorization_unsubtracted}, taking the advantage that the functions $f_{q}^{(u)},~S_{nn_h}$, and $J_{q}^{(u)}$ are all even function of $b$ as they depend on $b^2$, we further simplify the integration to be in the region $b\in (0, \infty)$. 

Finally, the well-known prefactor $\sigma_0$ is the leading-order (LO) partonic cross section for lepton-quark scattering
\begin{align}
\sigma_0
=
\frac{2 \alpha_{\mathrm{em}}^2}{s Q^2}
\frac{\hat{s}^2 + \hat{u}^2}{\hat{t}^2}
\, ,
\end{align}
where $\alpha_{\mathrm{em}}$ is the fine structure constant, $s$ is the center-of-mass energy squared of the incoming lepton and the proton beam, $Q^2$ represents the photon virtuality.
In the back-to-back lepton-hadron production region, the partonic Mandelstam variables $\hat{s},\ \hat{t}$ and $\hat{u}$ are connected to the Bjorken-$x$ and other kinematic variables:
\begin{align}
\hat{s}
& =
xs
\, , \label{e.shat} \\
\hat{t}
& =
-Q^2
=
- p_T^e e^{y_e} \sqrt{s}
\, , \label{e.that} \\
\hat{u}
& =
- x p_T^e e^{-y_e} \sqrt{s}
\, . \label{e.uhat}
\end{align}
For convenience, we also list here the Bjorken $x$ and inelasticity $y$ written in terms of other kinematical variables of interest:
\begin{align}
x
& =
\frac{p_T^e e^{y_e}}{\sqrt{s} - p_T^e e^{-y_e}}
\, , \\
\label{eq:x_Bj}
y
& =
1 - \frac{p_T^e}{\sqrt{s}} e^{-y_e}
=
\frac{Q^2}{xs}
\, ,
\end{align}
where we have used the momentum conservation relation $\hat{s} + \hat{t} + \hat{u} = 0$.

In the following subsections, we will identify all the components in the factorization theorem as given in~\cref{e.factorization_unsubtracted}.

\subsection{Quark distribution}
In this subsection, we provide a short overview of TMD quark distribution and discuss its expansion in terms of gluon dipole distribution in the small-$x$ limit. 

For the ``unsubtracted'' TMD quark distribution $f_{q}^{(u)} \pqty{x, b, \mu, \zeta/\nu^2}$, we have the Collins-Soper scale $\zeta$~\cite{Boussarie:2023izj,Ebert:2019okf,Collins:2011zzd} and a rapidity scale $\nu$~\cite{Chiu:2012ir}. The rapidity divergence in $f_{q}^{(u)}$ can be canceled by subtracting a square root of the standard soft function $S_{n \overline{n}} \pqty{{b},\mu, \nu}$ whose result at the next-to-leading order (NLO) is given by
\begin{align}
\allowdisplaybreaks
S_{n \overline{n}} \pqty{{b},\mu, \nu}
& =
1 -
\frac{\alpha_s C_F}{2 \pi} \Bigg[\ln[2](\frac{\mu^2}{\mu_b^2})
-
\frac{2}{\epsilon^2}+\frac{\pi^2}{6}
 \nnu
& +
2 \pqty{\frac{2}{\eta} + \ln(\frac{\nu^2}{\mu^2})}
\pqty{\frac{1}{\epsilon} + \ln(\frac{\mu^2}{\mu_b^2})}
\Bigg]
\, ,
\label{e.S_nnbar} 
\end{align}
where $\mu_b$ is defined as $\mu_b \equiv 2 e^{-\gamma_E}/b$.
It is worth noting that in this work we have applied the $4 - 2\epsilon$ space-time dimensions and the rapidity regulator $\eta$ \cite{Chiu:2012ir}.
As a consequence, we further defined the ``subtracted'' parton distribution $f_{q} \pqty{x, b, \mu, \zeta}$ without a rapidity divergence \cite{Collins:2011zzd}:
\begin{align}
f_{q} \pqty{x, b, \mu, \zeta}
=
f_{q}^{(u)} \pqty{x, b, \mu, \zeta/\nu^2}
\sqrt{S_{n\overline{n}} \pqty{{b}, \mu, \nu}}
\, .
\end{align}

TMD evolution for the ``subtracted'' TMD quark distribution is governed by two equations, the Collins-Soper evolution associated with the Collins-Soper scale $\zeta$~\cite{Boussarie:2023izj,Collins:2011zzd} and the renormalization group equation related to the scale $\mu$. They are given by
\begin{align}
\frac{\dd{}}{\dd{\ln\sqrt{\zeta}}}
\ln{f_q \pqty{x, b, \mu, \zeta}}
& =
K(b,\mu)
\, , \\
\frac{\dd{}}{\dd{\ln{\mu}}}
\ln{f_q \pqty{x, b, \mu, \zeta}}
& =
\gamma_\mu^q \bqty{\alpha_s(\mu), \zeta/\mu^2}
\, ,
\end{align}
where $K(b, \mu)$ denotes the Collins-Soper evolution kernel~\cite{Collins:2011zzd,Boussarie:2023izj,Moult:2022xzt,Duhr:2022yyp} and $\gamma_\mu^q \bqty{\alpha_s(\mu), \zeta/\mu^2}$ is given by:
\begin{align}
\gamma_\mu^q \bqty{\alpha_s(\mu),\frac{\zeta}{\mu^2}}
=
-\Gamma_{\mathrm{cusp}}^q \bqty{\alpha_s(\mu)}
\ln(\frac{\zeta}{\mu^2})
+
\gamma_\mu^q \bqty{\alpha_s(\mu)}
\, ,
\end{align}
where $\Gamma_{\mathrm{cusp}}^q$ and $\gamma_\mu^q$ are the cusp and non-cusp anomalous dimensions. They can be perturbatively expanded as:
\begin{align}
\Gamma_{\mathrm{cusp}}^q \bqty{\alpha_s(\mu)}
& =
\sum_{n=1} \Gamma_{n-1}^q \pqty{\frac{\alpha_s}{4 \pi}}^n
\, , \\
\gamma_\mu^q \bqty{\alpha_s(\mu)}
& =
\sum_{n=1} \gamma_n^q \pqty{\frac{\alpha_s}{4\pi}}^n
\, .
\end{align}
Solving the renormalization group equations on $\zeta$ and $\mu$ and taking into account the non-perturbative contribution at the large $b\gg 1/\Lambda_{\rm QCD}$ region, we obtain the TMD quark distribution as
\begin{align}
\allowdisplaybreaks
f_{q} \pqty{x, b, \mu, \zeta}
=&\, f_q \pqty{x, b,\mu_{b_*},\mu_{b_*}^2}
\exp[-S_{\rm NP}(b, Q_0, \zeta)]
\nnu
&\,\times \exp[-S_{\mathrm{pert}} (\mu, \mu_{b_*}, \zeta)] \,,
\label{e.TMD_standard}
\end{align}
where we evolve the TMD quark distribution $f_{q} \pqty{x, b, \mu_0, \zeta_0}$ at initial scales $(\mu_0, \zeta_0)$ to $f_{q} \pqty{x, b, \mu, \zeta}$ at final scales $(\mu, \zeta)$ and we have chosen the initial scales $\mu_0=\sqrt{\zeta_0} = \mu_{b_*}$. As usual, we define $\mu_{b_*} = 2e^{-\gamma_E}/b_*$ and $b_* = b/\sqrt{1+b^2/b_{\max}^2}$ with $b_{\max} = 1.5~\mathrm{GeV}^{-1}$ following the $b_*$-prescription in~\cite{Echevarria:2020hpy,Sun:2014dqm,Isaacson:2023iui,Collins:1984kg}. Here, $S_{\mathrm{pert}} \pqty{\mu, \mu_{b_*}, \zeta}$ is the perturbative Sudakov factor:
\begin{align}
\allowdisplaybreaks
S_{\mathrm{pert}} \pqty{\mu, \mu_{b_*} ,\zeta}
 =& 
- K \pqty{b_*, \mu_{b_*}} \ln(\frac{\sqrt{\zeta}}{\mu_{b_*}})
 \nnu
& -
\int_{\mu_{b_{*}}}^{\mu} \frac{\dd{\mu'}}{\mu'} \gamma_{\mu'}^q \bqty{\alpha_s(\mu'), \frac{\zeta}{\mu'^2}}
\, .\label{e.perturbative_Sudakov}
\end{align}
Throughout this paper, we will work at the next-to-leading logarithmic (NLL) level, where we have $K \pqty{b_*, \mu_{b_*}} = 0$ and we keep 
\begin{align}
    \Gamma_0^q = &\, 4C_F\,, \qquad
    \gamma_0^q = \, 6C_F\,,\\
    \Gamma_1^q = &\, 4C_F
\bqty{C_A\pqty{\frac{67}{18}-\frac{\pi^2}{6}} - \frac{10}{9} T_R n_f}\,.
\end{align}
On the other hand, $S_{\rm NP}(b, Q_0, \zeta)$ is a non-perturbative Sudakov factor for the TMD quark distribution, see e.g. Refs.~\cite{Echevarria:2020hpy,Sun:2014dqm}. In the conventional TMD approach~\cite{Boussarie:2023izj}, one would further express $f_q \pqty{x, b,\mu_{b_*},\mu_{b_*}^2}$ in terms of the collinear quark distribution functions through operator product expansion 
\begin{align}
\allowdisplaybreaks
    f_{q} \pqty{x, b, \mu_{b_*},\mu_{b_*}^2}
 =\,&
\sum_i \int _x^1 \frac{\dd{y}}{y}
C_{q \leftarrow i} \pqty{\frac{x}{y},b} 
f_i(y, \mu_{b_*})\,,
\end{align}
where $f_1^i \pqty{x, \mu_{b_*}}$ is the collinear quark distribution and $C_{q \leftarrow i}$ are the perturbatively calculable matching coefficients that can be found in e.g. Refs.~\cite{Aybat:2011zv, Collins:2011zzd, Kang:2015msa, Echevarria:2020hpy,Luo:2019szz,Luo:2020epw,Ebert:2020yqt}.

In this work, in order to explore the gluon saturation, following Refs.~\cite{Marquet:2009ca,Tong:2022zwp}, we expand this TMD quark distribution at the initial scale $\mu_0 = \sqrt{\zeta_0} = \mu_{b_*}$ in terms of the dipole gluon distribution at small $x$,
%%%%%%%%%%%%%%%%%%%%%%%%%%%%%%%%%%%%%%%%%%%%%%%%%%%%%%%
%%%%%%%%%%%%%%%%%%%%%%%%%%%%%%%%%%%%%%%%%%%%%%%%%%%%%%%
\begin{align}
&x f_q \pqty{x, b, \mu_{b_*}, \mu_{b_*}^2}\nnu
& =\,
\frac{N_c S_{\perp}}{8 \pi^4}
\int \dd{\epsilon_f^2} \dd[2]{\boldsymbol{r}}
\frac{\pqty{\bt + \rt} \vdot \rt}{\abs{\bt +\rt} \abs{\rt}} 
\epsilon_f^2 K_1 \pqty{\epsilon_f \abs{\bt +\rt}}
K_{1} \pqty{\epsilon_f \abs{\rt}}
\nnu
&\quad \,\times
\Big[1 + \mathcal{S}_x (\abs{\bt}) - \mathcal{S}_x \pqty{\abs{\bt + \rt}} - \mathcal{S}_x (\abs{\rt}) \Big]
\, ,
\label{e.small_x_PDFs}
\end{align} 
where $S_{\perp}$ is the averaged transverse area of the target hadron and $\mathcal{S}_x (r)$ represents the dipole scattering matrix with the dipole transverse size $r$. 
We consider two different models for $\mathcal{S}_x (r)$. 
The first is the Golec-Biernat-Wüsthoff (GBW) model~\cite{Golec-Biernat:1998zce, Golec-Biernat:1999qor} which can be written as:
\begin{align}
\mathcal{S}_x (r) = \exp \pqty{-\frac{r^2 \Qs^2(x)}{4}}
\, , \label{e.qs}
\end{align}
where the saturation scale $\Qs$ reads:
\begin{align}
\Qs^2(x) = 1~\mathrm{GeV}^{2} \times \left(\frac{x_0}{x}\right)^{\lambda}
\, .
\end{align}
The free parameters in this model are chosen as $\lambda = 0.29$, $x_0 = 3 \times 10^{-4}$ and $S_\perp =1/2 \times 23~\mathrm{mb}$ for proton targets following Ref.~\cite{Golec-Biernat:1998zce}.
The other model we consider is based on the McLerran-Venugopalan (MV) initial condition~\cite{McLerran:1993ka, McLerran:1993ni, McLerran:1994vd} which is then evolved with a running-coupling BK (rcBK) equation to smaller values in $x$.
Specifically, 
we use the MV$^e$ initial condition~\cite{Lappi:2013zma}:
\begin{align}
\mathcal{S}_x (r)
=
\exp\left[ -  \frac{r^2 Q_{s,0}^2}{4} \ln(\frac{1}{r \Lambda_\text{QCD}} + e_c \cdot e) \right]
\,,
\end{align}
and the rcBK equation:
 \begin{align}
 \allowdisplaybreaks
&\frac{\partial}{\partial \ln(1/x)} \mathcal{S}_x (\abs{\rt}) \nnu
& \hspace{-0.05cm}= \int \dd[2]{\rt'} \mathcal{K}(\rt, \rt') \qty\Big[ \mathcal{S}_x(\abs{ \rt'} )\mathcal{S}_x(\abs{\rt-\rt'} ) - \mathcal{S}_x(\abs{\rt}) ] \, .
 \end{align}
For the kernel $\mathcal{K}(\rt, \rt')$, we use the Balitsky prescription~\cite{Balitsky:2006wa}:
\begin{align}
\allowdisplaybreaks
\mathcal{K}(\rt, \rt')
& =
\frac{N_c \alpha_s(\rt^2)}{2 \pi^2} \left[ \frac{\rt^2}{\rt^{\prime 2} (\rt - \rt')^2} \right. \nnu
& \quad \left.+ \frac{1}{\rt^{\prime 2}} \qty( \frac{\alpha_s(\rt^{\prime 2})}{\alpha_s((\rt-\rt')^2)} -1 )\right.\nnu
& \quad \left. + \frac{1}{(\rt-\rt')^2} \qty( \frac{\alpha_s((\rt-\rt')^2)}{\alpha_s(\rt^{\prime 2})} -1 ) \right]\,,
\end{align}
with the coordinate-space running coupling
\begin{align}
    \alpha_s(r^2) = \frac{12 \pi}{(33-2 N_f) \ln( \frac{4 C^2}{r^2 \Lambda_\text{QCD}} )}\,.
\end{align}
We shall call this the rcBK model.
The values of the parameters for proton targets are taken from Ref.~\cite{Lappi:2013zma}, with the transverse size being $S_\perp = \sigma_0/2$ in terms of the parameters presented there. 
We also note that these parameter values are very close to the more recent ones in Ref.~\cite{Casuga:2023dcf} determined using Bayesian inference.

Finally, one has the following expression for TMD quark distributions in the CGC formalism, 
\begin{align}
f_{q} \pqty{x, b, \mu, \zeta}
=
f_q \pqty{x, b,\mu_{b_*},\mu_{b_*}^2}
\exp[-S_{\mathrm{pert}} (\mu, \mu_{b_*}, \zeta)]
\, ,
\label{e.fq_Spert}
\end{align}
with $f_q(x, b,\mu_{b_*},\mu_{b_*}^2)$ at the small-$x$ region provided in~\cref{e.small_x_PDFs} and the perturbative Sukadov factor given in~\cref{e.perturbative_Sudakov}. In comparison with the standard TMD quark distribution in~\cref{e.TMD_standard}, we ignore the non-perturbative Sudakov factor $S_{\rm NP}(b, Q_0, \zeta)$. This is because in principle the small-$x$ formula for the TMD quark distribution in~\cref{e.small_x_PDFs} has already contained the non-perturbative contribution in the large-$b$ region~\cite{Tong:2022zwp}.

\subsection{Hard and Soft functions}
The hard function $H \pqty{Q, \mu}$ with the renormalized expression at the one-loop is given by~\cite{Liu:2018trl, Arratia:2020nxw, Ellis:2010rwa}:
\begin{align}
\allowdisplaybreaks
H \pqty{Q, \mu}
=&\, 
1 + \frac{\alpha_s}{2\pi} C_F
\nnu
&  \hspace{-0.202cm}\times
\bqty{-\ln[2](\frac{\mu^2}{Q^2}) - 3 \ln(\frac{\mu^2}{Q^2}) - 8 +\frac{\pi^2}{6}} \, .
\end{align}
The natural scale for the hard function is given by $\mu\sim Q$.

On the other hand, the soft function $S_{nn_h} ({b},\mu, \nu)$ in DIS for TEEC at the NLO is given by:
\begin{align}
\allowdisplaybreaks
& S_{nn_h}({b},\mu, \nu)
=
1 -
\frac{\alpha_s C_F}{2\pi}
\Bigg[\ln[2](\frac{\mu^2}{\mu_b^2})
- \frac{2}{\epsilon^2} + \frac{\pi^2}{6}
 \nnu
& \quad +
2\pqty{\frac{2}{\eta}+\ln(\frac{\nu^2 \, n \cdot n_h/2}{\mu^2})}
\pqty{\frac{1}{\epsilon} + \ln(\frac{\mu^2}{\mu_b^2})}
\Bigg]
\, ,\label{e.S_nnh}
\end{align}
where $n \cdot n_h = 1-\tanh(y)$ with $y$ the rapidity of the final-state hadron.
This soft function can be related to the soft function for EEC in $e^+e^-$, namely the standard soft function $S_{n\overline{n}} \pqty{b,\mu,\nu}$ in \cref{e.S_nnbar} by \cite{Li:2020bub,Fang:2023thw}:
\begin{align}
S_{nn_h} \pqty{b,\mu,\nu}
=
S_{n\overline{n}} \pqty{b,\mu,\nu\sqrt{\frac{n \cdot n_h}{2}}}
\, .
\end{align}

\subsection{TEEC jet function and factorization}\label{sec:teec}

In \cref{e.factorization_unsubtracted}, the function denoted by $J_q^{(u)} \pqty{b, \mu, \zeta'/\nu^2}$ is the unsubtracted TEEC jet function~\cite{Moult:2018jzp} which is related to the ``unsubtracted'' transverse-momentum-dependent fragmentation functions (TMD FFs) via:
\begin{align}
J_q^{(u)} \pqty{b,\mu,\frac{\zeta'}{\nu^2}}
\equiv
\sum_h \int_0^1 \dd{z}
z \widetilde{D}^{(u)}_{1, h/{q}} \pqty{z,b,\mu,\frac{\zeta'}{\nu^2}}
\, ,
\end{align}
where the $\widetilde{D}^{(u)}_{1,h/{q}}(z,b,\mu,\zeta'/\nu^2)$ are the TMD FFs in the $b$-space. To simplify the notation, here we introduce the ``subtracted'' TMD FFs as:
\begin{align}
\label{e.subtracted_TMDFFs}
\allowdisplaybreaks
\widetilde{D}_{1, h/{q}} \pqty{z,b,\mu,\hat{\zeta}}
=
\widetilde{D}^{(u)}_{1, h/{q}} \pqty{z,b,\mu,\frac{\zeta'}{\nu^2}}
\frac{S_{nn_h}(b,\mu,\nu)}{\sqrt{S_{n\overline{n}}(b,\mu,\nu)}}
\, .
\end{align}
Using the results for the soft functions $S_{n\overline{n}} \pqty{b,\mu,\nu}$ and $S_{nn_h} \pqty{b,\mu,\nu}$ given in \cref{e.S_nnbar,e.S_nnh}, we find that the Collins-Soper scale for the ``subtracted'' TMD FFs $\widetilde{D}_{1, h/{q}}$ will be given by $\hat{\zeta} = \zeta'\pqty{n\cdot n_h/{2}}^2$. Note that in the rapidity regulator we adopt~\cite{Chiu:2012ir}, for TMD PDFs, the Collins-Soper scale is $\sqrt{\zeta}/2 = xP_A^+$, and for TMD FFs, one has $\sqrt{\zeta'}/2 = P_h^-/z$. Thus:
\begin{align}
\allowdisplaybreaks
&
Q^2
 =
-q^2
=
-\pqty{xP_A - \frac{P_h}{z}}^2=
2x\frac{P_A \cdot P_h}{z}
\nnu
& 
\quad=\sqrt{\zeta\zeta'}\,\frac{n\cdot n_h}{2}=\sqrt{\zeta\hat{\zeta}}
\, .
\end{align}
Namely, we find that $\zeta  \hat{\zeta} = Q^4$, and thus one can choose $\zeta = \hat{\zeta} = Q^2$ as a natural scale choice for the TMDs involved in the factorization formalism. 
Subsequently, the corresponding ``subtracted'' TEEC jet function $J_q \pqty{b,\mu,\hat \zeta}$ can be further written as:
\begin{align}
J_q \pqty{b,\mu,\hat{\zeta}}
\equiv
\sum_{h} \int_0^1 \dd{z} z \widetilde{D}_{1,h/q} \pqty{z,b,\mu,\hat{\zeta}}
\, .
\label{e.Jq}
\end{align}
The TMD FFs $\widetilde{D}_{1,h/q} \pqty{z,b,\mu,\hat{\zeta}}$ with QCD evolution is given by
\begin{align}
\allowdisplaybreaks
\widetilde{D}_{1,h/q} \pqty{z,b,\mu,\hat{\zeta}}
 =&
\sum_i \int_z^1 \frac{\dd{y}}{y}
C_{i\leftarrow q} \pqty{\frac{z}{y},b}D_{h/i} \pqty{y,\mu_{b_*}}
\nnu
& \times
\exp[-S_{\mathrm{pert}} \pqty{\mu,\mu_{b_*},\hat{\zeta}}]
 \nnu
& \times
\exp[-S_{\mathrm{NP}} \pqty{z,b,Q_0,\hat{\zeta}}]
\, ,\label{e.TMDFF}
\end{align}
where the matching coefficients $C_{i\leftarrow q}$ can be found in Refs.~\cite{Echevarria:2020hpy,Luo:2019szz,Luo:2020epw,Ebert:2020yqt}. The corresponding non-perturbative Sudakov factor is given by:
\begin{align}
S_{\mathrm{NP}} \pqty{z,b,Q_0,\hat{\zeta}}
=
\frac{g_2}{2} \ln(\frac{b}{b_*}) \ln(\frac{\sqrt{\hat{\zeta}}}{Q_0}) + g_1^D \frac{b^2}{z^2}
\, ,
\end{align}
with $g_2 = 0.84$ and $g_1^D = 0.042~\mathrm{GeV}^{2}$~\cite{Echevarria:2020hpy,Sun:2014dqm}.

Plugging~\cref{e.TMDFF} into~\eqref{e.Jq}, one thus obtains a general form for the TEEC jet function. If it were not for the $z$-dependence in the non-perturbative Sudakov term $\exp(-g_1^D {b^2}/{z^2})$ in~\cref{e.TMDFF}, one could decouple the $z$-integral in~\cref{e.Jq} with the $y$-integral in~\cref{e.TMDFF}:
\begin{align}
\allowdisplaybreaks
    &\sum_{h} \int_0^1 \dd{z} z \sum_i \int_z^1 \frac{\dd{y}}{y}
C_{i\leftarrow q} \pqty{\frac{z}{y},b}D_{h/i} \pqty{y,\mu_{b_*}}
\nnu
& =
\sum_{i} \sum_{h}\int_0^1 \dd{y} y\,D_{h/i}(y, \mu_{b_*})
\int_0^1 \dd{u} u\, C_{i\leftarrow q}\left(u, b\right)
\nonumber \\
& =
\sum_{i} \int_0^1 \dd{u} u\, C_{i\leftarrow q}\left(u, b\right)\,.
\end{align}
Here in the second line, we change the integration variable $u=z/y$, and in the third line, we apply the momentum sum rule, $\displaystyle \sum_{h} \int_0^1 \dd{z} z D_{h/q} \pqty{z,\mu_{b_*}} = 1$. This result is consistent with~\cite{Moult:2018jzp}. Unfortunately, the explicit $z$-dependence in the non-perturbative Sudakov factor $S_{\mathrm{NP}} \pqty{z,b,Q_0,\hat{\zeta}}$ makes the TEEC jet function more complicated. 

To proceed, we choose the coefficient function at the leading order $C_{i\leftarrow q} \pqty{z,b} = \delta_{iq}\delta(1-z)$ in \cref{e.TMDFF}, and thus the TEEC jet function $J_q$ in \cref{e.Jq} can be written as:
\begin{align}
\allowdisplaybreaks
  J_q \pqty{b,\mu,\hat{\zeta}}
=&\,
\sum_{h} \int_0^1 \dd{z} z D_{h/q} \pqty{z,\mu_{b_*}}
\exp(-g_1^D \frac{b^2}{z^2})
\nnu
&\,\times  \exp[-S_{\mathrm{pert}} \pqty{\mu,\mu_{b_*},\hat{\zeta}}]
\nnu
&\,\times \exp[-\frac{g_2}{2} \ln(\frac{b}{b_*}) \ln(\frac{\sqrt{\hat{\zeta}}}{Q_0})]
\,.  
\end{align}
Next, to prepare for the phenomenological study, we proceed by specifying a model for the TEEC jet function. Following~\cite{Li:2021txc}, we perform a fit to obtain a simple form for the $z$-integrated expression. Specifically, we define:
\begin{align}
\allowdisplaybreaks
& \sum_{h} \int_0^1 \dd{z} z D_{h/q} \pqty{z,\mu_{b_*}}
\exp(-g_1^D \frac{b^2}{z^2}) \nnu
&\equiv
\exp[-S_{\rm NP}^{\rm TEEC}(b)] \, ,
\end{align}
and use the DSS parameterization~\cite{deFlorian:2007aj} for all the hadrons $h=h^+, h^-, h^0$ in the fit. We find that the following functional form works very well:
\begin{align}
 S_{\rm NP}^{\rm TEEC}(b) 
= g_0^e \sqrt{b} + g_1^e b + g_2^e b^2\,.   
\end{align}
The fitted parameters are given by 
$g_0^e = 0.226$ GeV$^{1/2}$, $g_1^e = 0.463$ GeV, and $g_2^e = 0.033$ GeV$^2$. This fitted result is slightly different from what was obtained in Ref.~\cite{Li:2021txc}. 
Therefore, one has the TEEC jet function given by
\begin{align}
\allowdisplaybreaks
&J_q \pqty{b,\mu,\hat{\zeta}}
 =
\exp[-S_{\mathrm{pert}} \pqty{\mu,\mu_{b_*},\hat{\zeta}}]
\nnu
& \hspace{0.65cm}\times \exp[-\frac{g_2}{2} \ln(\frac{b}{b_*}) \ln(\frac{\sqrt{\hat{\zeta}}}{Q_0}) - S_{\rm NP}^{\rm TEEC}(b)]
 \, .
\end{align}

Eventually, one can write the factorization theorem in~\cref{e.factorization_unsubtracted} in terms of ``subtracted'' quark distributions and TEEC jet functions as:
\begin{widetext}
\begin{align}
\textbf{TEEC} = \frac{\dd{\sigma}}{\dd{\tau} \dd{y_e} \dd[2]{\boldsymbol{p}_T^e}}
=&\,
\sigma_0 H \pqty{Q, \mu} \sum_q e_q^2  \frac{p_T^e}{\sqrt{\tau}}
\int_0^\infty \frac{\dd{b}}{\pi} \cos(2b\sqrt{\tau} p_T^e) f_{q} \pqty{x, b, \mu, \zeta}
J_{q} \pqty{b,\mu, \hat{\zeta}}
\, . \label{e.factorization}
\end{align}
\end{widetext}
In the phenomenological section below, we choose the nominal scales $\mu=\sqrt{\zeta}=\sqrt{\hat\zeta}=Q$. As indicated in the introduction, when changing from $ep$ to $eA$ collisions, one takes nuclear modification effects into consideration and substitutes the saturation scale $Q_s$ in \cref{e.qs} by the nuclear saturation scale $Q_{s,A}^2\sim A^{1/3}\,Q_{s}^2$. More details about the numerical values of relevant parameters will be discussed in \cref{sec:pheno}.

\section{Phenomenology}\label{sec:pheno}
In this section, we make numerical predictions for the TEEC at the future EIC for both $ep$ and $eA$ collisions.

\begin{figure}[t]
\centering
\includegraphics[width=0.935\columnwidth]{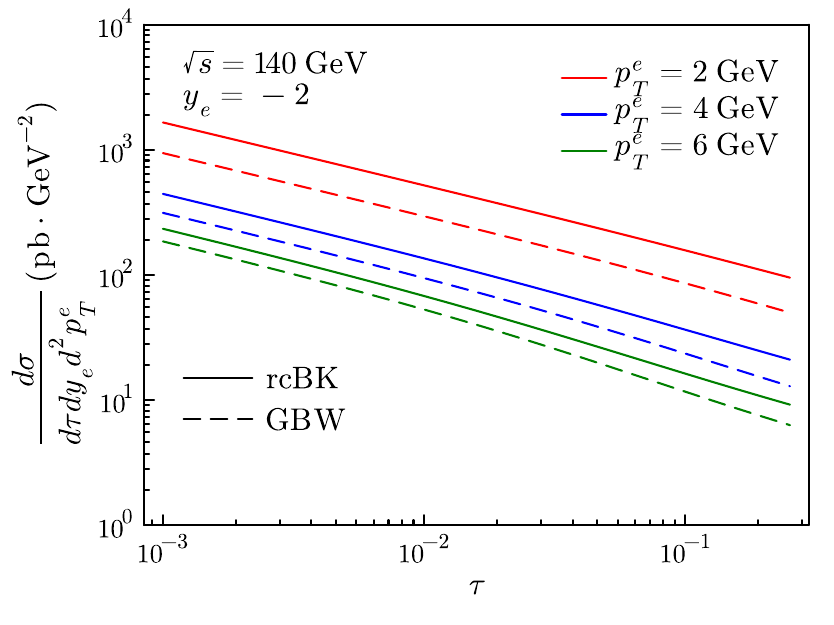}
\caption{The TEEC plotted as a function of $\tau$ for $e+p$ collisions at the future EIC. We choose the center-of-mass energy $\sqrt{s} = 140$ GeV and the lepton rapidity $y_e = -2$. The solid curves are from rcBK parameterizations while the dashed ones are based on the GBW model. The red, blue and green curves correspond to $p_T^e = 2, 4$ and 6 GeV, respectively.}
\label{fig:eic_ep}
\end{figure}

With the factorization of the TEEC jet function given in \cref{sec:teec}, we are now ready to perform numerical predictions for the TEEC at the future EIC.
We choose the highest center-of-mass energy $\sqrt{s} = 140$ GeV for electron-proton collisions.
We work in the frame where the proton is moving along the $+z$ direction, and the electron moves along the $-z$ direction.
In order to probe the small-$x$ region, we need to choose a proper lepton rapidity and transverse momentum.
As an example, we choose $y_e = -2$ and $p_T^e = 2,\, 4,\, 6$ GeV. This corresponds to the probed $x$ values between $2 \times 10^{-3}$ and $8.5 \times 10^{-3}$.
In \cref{fig:eic_ep}, we plot the TEEC as a function of $\tau$ for these three different $p_T^e$ values. The solid curves are from the rcBK parameterizations while the dashed ones are based on the GBW model. The red curves are for $p_T^e = 2$ GeV, the blue ones are for $p_T^e = 4$ GeV, and the green ones are for $p_T^e=6$ GeV. We find that the numerical results based on the rcBK and the GBW model for the TEEC observables can differ by a factor of two, especially at the small $p_T^e = 2$ GeV, indicating the TEEC at the EIC can be a good observable constraining the dipole gluon distribution. 

To study the nuclear modification in $e+A$ collisions in comparison with the $e+p$ scatterings, we define the nuclear modification factor $R_A$ as follows:
\begin{align}
\label{eq:RA}
R_{A} =
\frac{1}{A} \left.
\frac{\dd{\sigma_{eA}}}{\dd{\tau} \dd{y_e} \dd[2]{\boldsymbol{p}_T^e}}
\right/
\frac{\dd{\sigma_{ep}}}{\dd{\tau} \dd{y_e} \dd[2]{\boldsymbol{p}_T^e}}
\, ,
\end{align}
where $A$ is the atomic mass of the nuclear target. Below, we choose the gold nucleus with $A=197$. To go from the proton to the nucleus beam, we change the proton saturation scale to the nuclear saturation scale $Q_{s,A} \pqty{x}$ or $Q_{s,0,A}^2$ for the GBW and the rcBK models, respectively:
\begin{figure}[b]
\centering
\includegraphics[width=0.9\columnwidth]{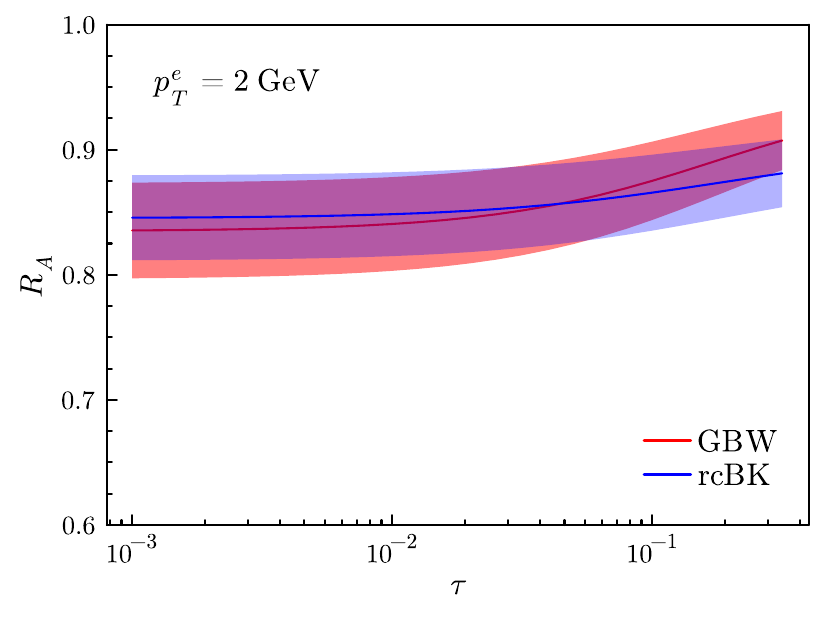}
\\
\includegraphics[width=0.9\columnwidth]{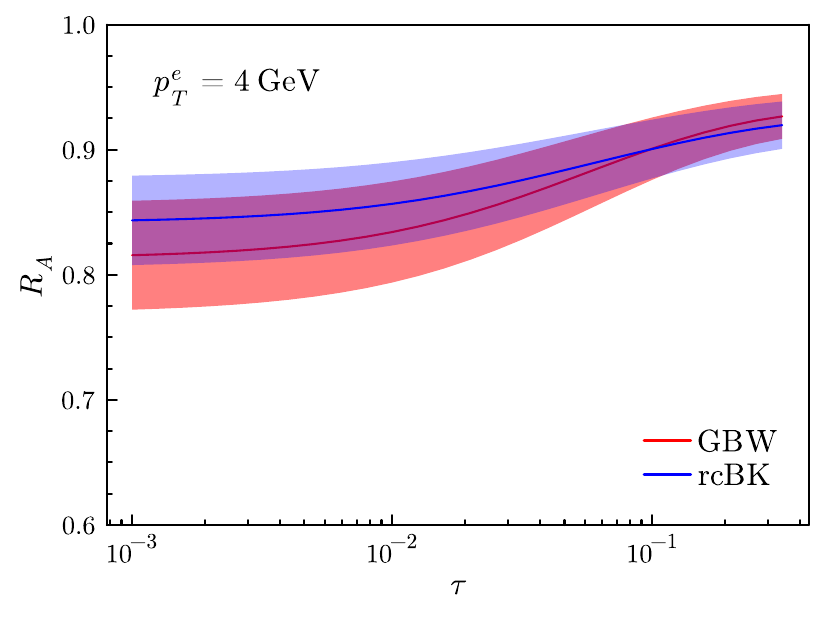}
\\
\includegraphics[width=0.9\columnwidth]{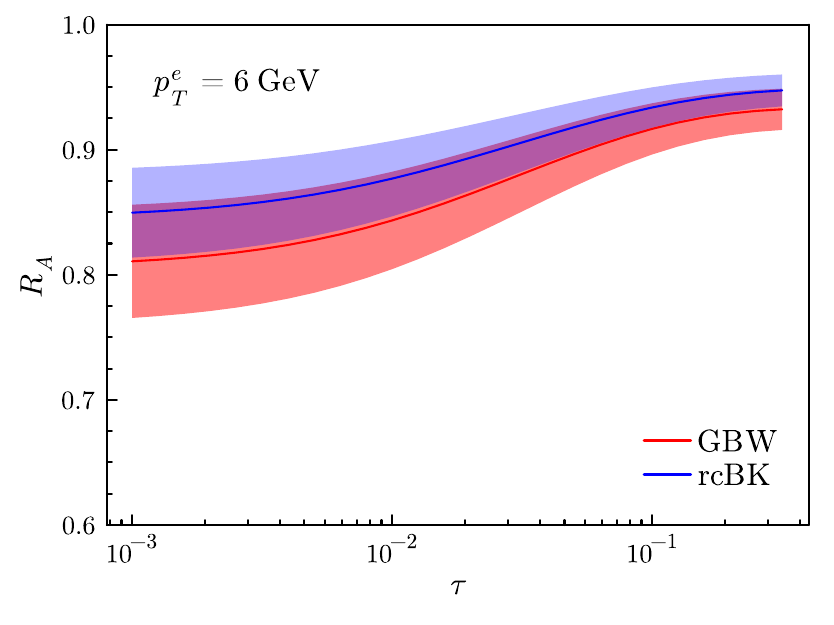}
\caption{Nuclear modification factor $R_A$ from \cref{eq:RA} is plotted as a function of $\tau$ for $p_T^e=2$ GeV (top panel), 4 GeV (middle panel), and 6 GeV (bottom panel). We choose $\sqrt{s}=140$ GeV and lepton rapidity $y_e = -2$. The red bands are for the GBW model, and the blue bands are for the rcBK calculations.}
\label{fig:RA_pT}
\end{figure}
\begin{align}
Q_{s,A}^2 \pqty{x} = c \, A^{1/3} \, Q_{s}^2 \pqty{x}
\, , \quad
Q_{s,0,A}^2 = c \, A^{1/3} \, Q_{s,0}^2
\, ,
\end{align}
where $Q_{s} \pqty{x}$ and $Q_{s,0}$ are the proton saturation scales for the GBW and the rcBK models, respectively.
The parameter $c$ is chosen in the range $0.5 < c < 1.0$~\cite{Dusling:2009ni, Tong:2022zwp}.
Correspondingly, we also change the active nuclear transverse area $S_{\perp} \to S_{\perp, A} = 1/c \times A^{2/3} \, S_{\perp}$.
Having the same scaling constant $c$ for both the saturation scale and the transverse area can be motivated by the smooth nucleus approach~\cite{Kowalski:2003hm}, where in the dilute region we can integrate over the impact parameter and write $Q_{s,A}^2 \pqty{x} = A S_{\perp} / S_{\perp, A} \times Q_s^2 \pqty{x}$ for the GBW model or $Q_{s,0,A}^2 = A S_{\perp} / S_{\perp, A} \times  Q_{s,0}^2$ for the rcBK model. In the future, we plan to implement the impact parameter dependence~\cite{Kowalski:2003hm,Mantysaari:2018nng,Deganutti:2023qct} directly inside the saturation formalism and thus provide more accurate predictions.

In \cref{fig:RA_pT}, we plot the nuclear modification factor $R_A$ as a function of $\tau$ for $p_T^e=2$ GeV (top panel), 4~GeV (middle panel), and 6~GeV (bottom panel). We choose $\sqrt{s}=140$ GeV and lepton rapidity $y_e = -2$. The bands correspond to the uncertainty in the parameter $0.5 < c < 1.0$. The red bands are for the GBW model, and the blue bands are for the rcBK calculations. It shows that nuclear modifications on the order of $15\%-20\%$ can be expected in the small $\tau$ region, for both the rcBK and the GBW model. On the other hand, the nuclear modification factor starts to approach 1 as the $\tau$ value increases. Such behavior is a manifestation of the $\cos(2b\sqrt{\tau} p_T^e)$ modulation in \cref{e.factorization}. In the large $\tau$ region, the integration is dominated by the small-$b$ region where the dipole size is small and thus the saturation effect is less important and one expects $R_A\to 1$. On the other hand, in the small $\tau$ region, one would receive more contribution from the larger dipole size (large $b$ region) and correspondingly stronger nuclear modification. This indicates that the TEEC is a good observable for gluon saturation.

\section{Conclusions}\label{sec:conclusion}
In this paper, we explore the transverse energy-energy correlators in the small-$x$ regime for the future EIC. For the production of electron-hadron pairs in the back-to-back region in the transverse plane where the azimuthal angle difference $\phi\to \pi$ between the final-state lepton and the hadron, we provide a factorization theorem that incorporates the gluon saturation effects. We present numerical results for TEEC in both $e+p$ and $e+A$ collisions, alongside evaluations of the nuclear modification factor $R_A$. We find that the TEEC observables in $e+p$ collisions are significantly influenced by different models of the dipole gluon distribution, emphasizing the potential of TEEC at the EIC as a robust observable for constraining the dipole gluon distribution in the small-$x$ region. We introduce the variable $\tau = (1+\cos\phi)/2$, and our results indicate that the nuclear modification factor $R_A$ for TEEC exhibits a suppression in the range of $15\%-20\%$ in the small $\tau$ region. Conversely, as $\tau$ increases, $R_A$ tends toward unity. This trend aligns with expectations, as larger $\tau$ values correspond to smaller dipole sizes being probed by TEEC, resulting in reduced nuclear modifications.

The demonstrated potential of measuring TEEC at the EIC underscores its importance in improving our understanding of gluon saturation and nuclear modifications. As the EIC becomes operational, we anticipate that the insights gained from TEEC measurements will play a pivotal role in refining our understanding of the fundamental aspects of strong interaction physics.\\
\break

\section*{Acknowledgments}
Z.K. and J.P. are supported by the National Science Foundation under grant No. PHY-1945471. F.Z. is supported by U.S. Department of Energy, Office of Science, Office of Nuclear Physics under grant Contract Number DESC0011090 and U.S. Department of Energy, Office of Science, National Quantum Information Science Research Centers, Co-design Center for Quantum Advantage (C2QA) under Contract No. DESC0012704. Y.Z. is supported by the Guangdong Major Project of Basic and Applied Basic Research No.~2020B0301030008, and the National Natural Science Foundation of China under Grants No.~12022512 and No.~12035007. This work is also supported by the U.S. Department of Energy, Office of Science, Office of Nuclear Physics, within the framework of the Saturated Glue (SURGE) Topical Theory Collaboration.

\bibliographystyle{JHEP-2modlong.bst}
\bibliography{main.bib}
\end{document}